\documentclass[12pt]{article}
\begin{document}
\pagestyle{plain}
\setcounter{page}{1}
\baselineskip16pt

\begin{titlepage}

\begin{flushright}
PUPT-1574\\
hep-th/9510200
\end{flushright}
\vspace{20 mm}

\begin{center}
{\huge The Size of $p$-Branes}

\vspace{5mm}

\end{center}

\vspace{10 mm}

\begin{center}
{\large Igor R.\ Klebanov and L\'arus Thorlacius }

\vspace{3mm}

Joseph Henry Laboratories\\
Princeton University\\
Princeton, New Jersey 08544

\end{center}

\vspace{2cm}

\begin{center}
{\large Abstract}
\end{center}

We obtain form factors for scattering of gravitons and anti-symmetric
tensor particles off Dirichlet $p$-branes in Type II superstring theory.
As expected, the form factor of the $-1$-brane (the D-instanton)
exhibits point-like behavior: in fact it is saturated by the 
dilaton tadpole graph. In contrast, $p$-branes with $p>-1$ acquire
size of order the string scale due to quantum effects and exhibit
Regge behavior.  We find their leading form factors in closed form and
show that they contain an infinite sequence of poles associated with
$p$-brane excitations.  Finally, we argue that the $p$-brane form
factors for scattering of R-R bosons will have the same stringy features
found with NS-NS states.

\vspace{2cm}
\begin{flushleft}
October 1995
\end{flushleft}
\end{titlepage}
\newpage
\renewcommand{\baselinestretch}{1.1}  


\renewcommand{\epsilon}{\varepsilon}
\newcommand{\SU}[1]{\mbox{${\rm SU}(#1)$}}
\newcommand{\WZ}{{\rm WZ}}
\newcommand{\QCD}{{\rm QCD}}
\newcommand{\disk}{{\rm D}}
\newcommand{\grad}{\nabla}
\newcommand{\MeV}{{\rm MeV}}
\newcommand{\rightvbar}[1]{\left. #1 \right|}
\newcommand{\tr}{\mathop{\rm tr}}
\newcommand{\half}{{1\over 2}}
\newcommand{\sym}{\mathop{\rm sym}}
\newcommand{\id}{1}
\newcommand{\mtxIIxI}[2]{\parens{
   \begin{array}{c}
      #1 \\
      #2
   \end{array}
}}
\newcommand{\mtxIIxII}[4]{\parens{
   \begin{array}{cc}
      #1 & #2 \\
      #3 & #4
   \end{array}
}}

\newcommand{\sA}{{\cal A}}
\newcommand{\sB}{{\cal B}}
\newcommand{\sD}{{\cal D}}
\newcommand{\sL}{{\cal L}}
\newcommand{\sM}{{\cal M}}

\newcommand{\va}{\vec{a}}
\newcommand{\vb}{\vec{b}}
\newcommand{\vn}{\vec{n}}
\newcommand{\vv}{\vec{v}}
\newcommand{\vw}{\vec{w}}
\newcommand{\vx}{\vec{x}}
\newcommand{\vJ}{\vec{J}}
\newcommand{\valpha}{\vec{\alpha}}
\newcommand{\dalpha}{\dot{\valpha}}
\newcommand{\vtau}{\vec{\tau}}

\newcommand{\bc}{\bar{c}}
\newcommand{\bs}{\bar{s}}
\newcommand{\bomega}{\bar{\omega}}

\newcommand{\unitr}{\widehat{r}}
\newcommand{\unittheta}{\widehat{\theta}}
\newcommand{\unitphi}{\widehat{\phi}}

\newcommand{\dr}{{\rm\bf d}r}
\newcommand{\dtheta}{{\rm\bf d}\theta}
\newcommand{\dphi}{{\rm\bf d}\phi}
\newcommand{\dt}{{\rm\bf d}t}

String theory appears to contain more than just strings.
One lesson of the exciting recent developments is that higher
dimensional structures, so called $p$-branes, play a prominent
role in the strongly coupled dynamics
\cite{duff,chpt,witten,strominger}.
The $p$-branes arise as non-perturbative solutions of the
low-energy effective field theory of closed string theories
\cite{mdjl,chs} and are a crucial ingredient in various
string-string dualities \cite{schwarz,sen,jhas}.
In a seminal paper, Polchinski has proposed a realization of
$p$-branes in terms of mixed Dirichlet and Neumann boundary
conditions in Type-II superstring theory \cite{polchinski}.
Such a worldsheet approach allows one to study these extended
objects using well understood tools of two-dimensional quantum
field theory. This insight has already produced rapid development in the
field \cite{ed,li}.

The issue we wish to address here is that of the physical size
of $p$-branes.  This question is of special interest in light of
Shenker's recent suggestion \cite{shenker}, that certain
non-perturbative states
carrying RR charge may introduce a new dynamical length scale into
string theory, which is much shorter than the string scale if the
theory is weakly coupled. Our main conclusion is that,
at least as far as their gravitational properties are concerned,
the Dirichlet $p$-branes with $p>-1$ acquire a size of the order of the
string scale through quantum effects (the $-1$-brane is a special
case, as we shall see below).

We study the size of the Dirichlet $p$-branes by scattering off them
massless string states, whose vertex operators are
$$
V(z, \bar z) = \epsilon_{ij} (\partial X^i+ik\cdot \psi \psi^i)
( \bar\partial X^j +ik\cdot \bar\psi \bar\psi^i)
e^{ik\cdot X} \,,
$$
with $\epsilon_{ij}$ symmetric for gravitons and dilatons,
and anti-symmetric for
anti-symmetric tensor particles.  The resulting form factor provides
a measure of the effective thickness of the $p$-brane as seen
by stringy probes.  In this paper we restrict our attention to
amplitudes involving NS-NS sector string states, whereas one would
employ R-R sector photons to measure the R-R charge radius of a
Dirichlet $p$-brane.  There are technical reasons to believe our
result, that $p>-1$ Dirichlet-branes are typically of string scale
thickness, will extend to their R-R charge as well.  If this is the
case, one may have to look elsewhere for objects much smaller than
strings.  We will comment on this issue below and hope to present a
calculation of the scattering of R-R states off p-branes
at a later date.

The leading contribution to the $p$-brane form factor may be read off
from the two-point function of closed-string vertex operators on a disk
with the appropriate mixed Dirichlet and Neumann boundary conditions.
This process may be thought of as follows: an open string with both
end-points glued to the $p$-brane is created from vacuum, it absorbs
and emits a graviton (or anti-symmetric tensor particle) and is
subsequently annihilated. Since the open strings with end-points glued
to the $p$-brane describe excitations of the $p$-brane itself \cite{dlp}
there is another physical picture of the same process: a particle
collides with a $p$-brane creating an excitation propagating along
its world volume and the $p$-brane subsequently returns to its ground
state by emitting another particle.

Let us first consider this calculation for the simplest Dirichlet brane
with $p=-1$, also known as the D-instanton\cite{Green,jp2}. 
Since this object has no world
volume one might expect the scattering process described above to
degenerate and we shall see that this is indeed the case.
In discussing the $-1$-brane
one imposes Dirichlet boundary conditions on all ten superstring
coordinates. The required Green function on a disk of radius equal to 1
is given by
\begin{equation}
 \langle X^\mu (z, \bar z) X^\nu (w, \bar w)\rangle=
\eta^{\mu\nu}\bigl (-2\ln |z-w|+2 \ln |1-z\bar w|\bigr ) \,.
\label{dbos}\end{equation}
For the fermions we correspondingly take
\begin{equation}
\langle\psi^\mu(z) \psi^\nu (w)\rangle ={\eta^{\mu\nu} \over z-w}\ ,
\qquad
\langle\psi^\mu(z) \bar\psi^\nu (\bar w)\rangle
 ={i\eta^{\mu\nu} \over 1- z\bar w}\ .
\label{dferm}\end{equation}
One needs to evaluate the correlation function of two vertex operators,
one with polarization $\epsilon^1$ and momentum $k$ and the other
with polarization $\epsilon^2$ and momentum $p$.  The usual kinematics
of massless closed-string states applies:
$\epsilon^1_{\mu\nu}k^\nu=0=\epsilon^2_{\mu\nu}p^\nu$,
$\epsilon_\mu^{1\>\mu}=0=\epsilon_\mu^{2\>\mu}$.
We gauge fix the residual symmetry of the disk amplitude by integrating
only over the location of one of the vertex operators, placing the other
at the origin.  This, in fact, overcounts by a factor of $2\pi$ (from
the angular integration) which is easily divided out.
The result may be expressed as
\begin{equation}
\int d^{10} q e^{iq\cdot Y} F(q)
\label{formfactor}
\end{equation}
where $q=p+k$, and $Y$ is the position of the D-instanton.

In bosonic string theory the form factor $F(q)$ was first calculated by
Green and Wai \cite{mgpw} using string field theory techniques.
They found a simple answer with poles only at $q^2=-2, 0, 2$. This
expression exhibited a power law behavior at large $q^2$ indicative of
a point-like nature of the D-instanton.  It thus seems that string theory
contains objects which, at least in perturbation theory, are point-like
events.

We have calculated the D-instanton form factor in the Type-II superstring
theory and found it to be even simpler than in the bosonic theory.
As one might have expected, all poles at $q^2 =\pm 2$ cancel out.
In fact, for gravitons the entire two-point function vanishes!
For two anti-symmetric tensor particles the form factor
turns out to be
\begin{equation}
F_B (q) = \epsilon^1_{\mu\nu} \epsilon^2_{\nu\mu}
+ 4 \epsilon^1_{\mu\lambda} \epsilon^2_{\nu\lambda}
{q^\mu q^\nu \over q^2 } \,.
\label{dinst}
\end{equation}
This expression has a purely field theoretic interpretation:
it is due to a dilaton emitted by the $H^2 e^{-2\phi}$
vertex of the low-energy effective field theory (where $H_{ijk}$ is
the anti-symmetric tensor field strength)
and absorbed by a dilaton tadpole created by the D-instanton.
This dilaton tadpole does not affect the graviton two-point function
because in the physical ``Einstein frame'' there is no tree level
vertex coupling two on-shell gravitons to a dilaton. 

We conclude that in the Type-II superstring theory a
D-instanton is indeed  point-like, its form factors being
miraculously saturated by the field theoretic dilaton tadpole graph. 
While a fixed D-instanton produces such a tadpole, in the complete
theory one has to integrate over the collective coordinates.
In addition to the space-time 
position $X^\mu$ one finds its superpartner \cite{Green},
a Majorana-Weyl coordinate $\theta^a$, $a=1, 2, \ldots, 16$.
Integration over $\theta^a$ sets the dilaton tadpole and
the associated cosmological constant to zero.\footnote{We thank E. Witten
for pointing this out.} 

We now move on to consider Dirichlet $p$-branes with $p>-1$.
These objects are more physically familiar than the D-instanton:
the $0$-brane is a stringy description of a solitonic particle,
the $1$-brane describes a solitonic string, {\it etc}.
The gravitational form factors of these objects
turn out to have exponential rather than power law decay at
high energy, and do not exhibit any obvious point-like structure.
This behavior is very different from the D-instanton case.

To see what makes $p>-1$ so different from $p=-1$ let us consider
the $0$-brane. Here the spatial coordinates
$X^i$, with $i=1, 2, \ldots, 9$,
and the corresponding fermionic fields, satisfy Dirichlet boundary
conditions on the disk with Green functions given in
(\ref{dbos})-(\ref{dferm}).
The time coordinate, however, has Neumann boundary conditions,
so this time the Green functions for $X^0$ and $\psi^0$ are
\begin{equation}
 \langle X^0 (z, \bar z) X^0 (w, \bar w)\rangle=
2\ln |z-w|+2 \ln |1-{1\over z\bar w}|
\label{dgreen}
\end{equation}
\begin{equation}
\langle\psi^0(z) \psi^0 (w)\rangle = {-1\over z-w}\ ,
\qquad
\langle\psi^0(z) \bar\psi^0 (\bar w)\rangle = {i\over 1- z\bar w}\ .
\label{dfgreen}
\end{equation}
The change in sign of $\langle\psi^0(z) \bar\psi^0 (\bar w)\rangle$
compared to (\ref{dferm}) is required by worldsheet supersymmetry.
The asymmetry between the time and the spatial directions gives
rise to some interesting effects. For example, the normal ordering of
$e^{ik\cdot X}$ introduces a factor $|1-z\bar z|^{-2 k_0^2}$
into the correlation function. This factor, which was absent for the
D-instanton, leads to singular behavior near the edge of the disk.
The interplay between this singularity and the one arising from the
collision of the two vertex operators makes the amplitude stringy
rather than field theoretic.

Let us break up the momenta into spatial and time components,
\begin{equation}
 p= (p_0, \vec p) \ ,\qquad k= (k_0, \vec k) \ .
\label{momenta}
\end{equation}
For simplicity, we choose $\epsilon^1_{ij}$
and $\epsilon^2_{ij}$ to have non-zero components only in the
spatial directions ($1, 2, \ldots, 9$). Calculations with
general polarizations are somewhat more involved and this
restricted kinematics is sufficient to get at the physical result
we are interested in.  Since the energy is conserved,
$p_0=-k_0$, the form factor is a function of two quantities,
$k_0^2$ and $\vec k\cdot \vec p$.
The on-shell conditions are
\begin{equation}
 \vec p^2 = \vec k^2 =k_0^2\ , \qquad \epsilon^1_{ii}=\epsilon^2_{ii}=0\ ,
\qquad k^i \epsilon^1_{ij} = p^i \epsilon^2_{ij} =0 \ .
\label{onshell}
\end{equation}

As a warm up, let us perform the calculation in bosonic
string theory.
In calculating the term $\sim \epsilon^1_{ij}\epsilon^2_{ij}$
of the two-graviton amplitude we fix one
of the graviton vertex operators at the center of the disk
and find the following integral in the variable $x=r^2$,
$$
{1\over 2} \epsilon^1_{ij}\epsilon^2_{ij} \int_0^1 dx
\left [{1\over x^2} +1
\right ] (1-x)^{-2 k_0^2} x^{\vec k\cdot \vec p+ k_0^2} \ .
$$
Each of the necessary integrals yields a beta function as is common
in open string calculations.
This is not too surprising since
in the $s$ channel this process is mediated by an open string
sliding along the $p$-brane world volume.
The complete answer for the
gravitational form factor turns out to be
\begin{eqnarray}
& F_g^{bosonic} = {\Gamma(1-2k_0^2)
\Gamma (\vec k\cdot \vec p +k_0^2-1)\over
\Gamma (\vec k\cdot \vec p-k_0^2+2)}
\bigg [ \epsilon^1_{ij}\epsilon^2_{ij}
(k_0^4 -k_0^2 +(\vec k\cdot \vec p)^2)
 -2 \epsilon^1_{ij}
\epsilon^2_{il} p^j k^l (1-2k_0^2) \vec k\cdot \vec p
& \cr &
+ p^i\epsilon^1_{ij}p^j k^s\epsilon^2_{sl}k^l (1-k_0^2) (1-2k_0^2)
\bigg ] &
\end{eqnarray}
The anti-symmetric tensor form factor is
\begin{equation}
F_B^{bosonic} = {\Gamma(2-2k_0^2)
\Gamma (\vec k\cdot \vec p +k_0^2-1)\over \Gamma (\vec k\cdot \vec p-k_0^2+2)}
\bigg [ \vec k\cdot \vec p \epsilon^1_{ij}\epsilon^2_{ij}
- 2 \epsilon^1_{ij}\epsilon^2_{il} p^j k^l (1- k_0^2)
\bigg ]
\label{bosb}\end{equation}
This expression can be cast in a manifestly gauge invariant form,
\begin{equation}
F_B^{bosonic} = {\Gamma(2-2k_0^2)
\Gamma (\vec k\cdot \vec p +k_0^2-1)\over \Gamma (\vec k\cdot \vec p-k_0^2+2)}
\bigg [ \vec k\cdot \vec p H^1_{0ij}(k) H^2_{0ij}(p)
- {1- k_0^2\over 3} H^1_{ijl} (k) H^2_{ijl} (p) 
\bigg ]\ ,
\end{equation}
where
\begin{equation}
H^1_{\alpha \beta \gamma} (k) = i( k_\alpha \epsilon^1_{\beta \gamma}
+ k_\beta \epsilon^1_{\gamma \alpha}+ k_\gamma \epsilon^1_{\alpha \beta})
\ .\end{equation}

In proceeding to the type-II string one finds, as usual, that the
calculations are longer, but the answers are shorter.
We find the following expression for the gravitational form factor,
\begin{equation}
F_g={1\over 2}\epsilon^1_{ij}\epsilon^2_{ij} \int_0^1 dx
\left [{(1{-}k_0^2{-}\vec k\cdot \vec p)^2\over x^2}
+2 { k_0^4{-}(\vec k\cdot \vec p)^2\over x}
+ (1{-}k_0^2{+}\vec k\cdot \vec p)^2
\right ]
x^{\vec k\cdot \vec p+ k_0^2}
(1-x)^{-2 k_0^2} \,.
\label{simptwo}\end{equation}
Each of the necessary integrals is again a beta function and can be
explicitly evaluated to give the complete form factor,
\begin{equation}
F_g=\epsilon^1_{ij}\epsilon^2_{ij}k_0^2
{\Gamma (1-2k_0^2) \Gamma (\vec k\cdot \vec p +k_0^2)
\over \Gamma (\vec k\cdot \vec p-k_0^2+1)} \ .
\label{gravf}\end{equation}
We note that the terms involving contractions between the polarization
tensors and the transverse momenta have cancelled in this form factor.
The corresponding calculation for two anti-symmetric tensor particles
also yields a simple result,
\begin{equation}
F_B=\bigg [ 2 \epsilon^1_{ij}\epsilon^2_{il} p^j k^l 
-\epsilon^1_{ij}\epsilon^2_{ij}\vec k\cdot \vec p\bigg ]
{\Gamma (1-2k_0^2) \Gamma (\vec k\cdot \vec p +k_0^2)\over
\Gamma (\vec k\cdot \vec p-k_0^2+1)} \ ,
\label{genant}\end{equation}
whose manifestly gauge invariant form is
\begin{equation}
F_B= 
{1\over 3} H^1_{ijl} (k) H^2_{ijl} (p) 
{\Gamma (1-2k_0^2) \Gamma (\vec k\cdot \vec p +k_0^2)\over
\Gamma (\vec k\cdot \vec p-k_0^2+1)} \ .
\label{gaugeant}\end{equation}

Equations (\ref{gravf})-(\ref{gaugeant}) have structure typically found
in string theoretic four-point functions.  In the physical region there
is an infinite sequence of poles at $2 k_0^2 =1, 2, 3, \ldots$.
The positions of these poles coincide with the excitation energies
of an open string
whose ends are attached to the particle ($0$-brane). From this point of
view the infinite energy spectrum is perfectly natural.
We should emphasize, however, that it is somewhat unusual to find a
``particle'' with an infinite spectrum of excitations. Thus, the string
solitons which are described by the $0$-branes do not behave like
typical field theoretic particles. It appears that in string theory all
objects, even those that at low energies are best described as
particles, acquire stringy properties.

The difference between the D-instanton and the other $p$-branes could
be predicted on purely kinematical grounds. For the D-instanton, the
$\epsilon^1_{ij}\epsilon^2_{ij}$ term in the form factor is
{\it a priori} a function of $t=-(p+k)^2$ only. For the $0$-brane it
is a function of two variables,
$t=-(p+k)^2$ and the conserved quantity $s=k_0^2$.
This is what makes the Veneziano-type formulae
with the interplay between the $s$ and $t$ channels, possible.
For instance,
the graviton form factor (\ref{gravf}) may be rewritten as
\begin{equation}
F_g=s\, \epsilon^1_{ij}\epsilon^2_{ij}
{\Gamma (1-2s) \Gamma (-t/2)\over \Gamma (1-2s-t/2)}
\label{gravnew}\end{equation}
Here we find a somewhat unusual situation: the poles in the $s$-channel
correspond to an open string glued to the $p$-brane,
while the poles in the $t$-channel
correspond to closed strings created
in a collision of two massless particles
and absorbed by the $p$-brane. In the $u$-channel the process
appears essentially the same as in the $s$-channel and is also mediated
by the $p$-brane excitations.

As we proceed to $p$-branes with $p>0$, the number of variables in the
form factor
remains equal to two. For the $1$-brane, for example, in addition to
$t=-(p+k)^2$ we have the invariant conserved quantity
$s=k_0^2 -k_1^2$.
For gravitons polarized transversely to the
$1$-brane the form factor is still given by (\ref{gravnew}).
It is straightforward to generalize the calculation
of form factors to $1$-branes, $2$-branes, {\it etc}.
For gravitons and anti-symmetric tensor
particles polarized transversely to the $p$-brane world volume,
the results are given by (\ref{gravf})-(\ref{genant})
with $k_0^2$ replaced by an appropriate conserved
quantity:
$k_0^2\rightarrow k_0^2 - k_1^2$ for $1$-branes;
$k_0^2\rightarrow k_0^2 - k_1^2- k_2^2$
for $2$-branes; {\it etc}.
This demonstrates a remarkable universality in the properties of
the Dirichlet $p$-branes with $p>-1$: to gravitons or antisymmetric tensor
particles polarized transversely to the $p$-brane 
world volume all $p$-branes appear essentially the same.
In particular, for gravitons polarized transversely to the
$1$-brane (string) the form factor is still given by (\ref{gravnew}).
This remarkably simple expression for scattering off a long string
may serve as a guide to similar calculations where the long string
is described by a soliton.
Our conclusion from all of the above is that, as far as their
gravitational properties are concerned, all $p$-branes with $p>-1$
behave in a stringy, rather than point-like manner.

In this paper we presented detailed results for scattering of
NS-NS closed string states off the Dirichlet $p$-branes of type-II
superstring theory. An equally interesting exercise is to scatter
R-R closed string states off $p$-branes. While this is technically
more difficult, we may anticipate the structure of the result
simply by noting that R-R vertex operators also contain a
factor $e^{ik\cdot X}$.
The normal ordering of this factor introduces $|1- r^2|^{-2 s}$
into the integrand, which is singular near the edge of the disk
($s= k_0^2- \sum_{i=1}^p k_i^2 $ is one of the two kinematical
invariants for the scattering process). Thus, we anticipate that
the result will have a stringy structure similar to
(\ref{gravf})-(\ref{gravnew}), with an interplay between
the $s$ and $t$ channels. Another argument in favor of this
is the fact that the R-R scattering amplitudes are
related to the NS-NS scattering amplitudes by 
space-time supersymmetry. Our tentative conclusion is therefore
that quantum fluctuations smear the R-R charge of the $p$-branes
over the string scale.  The R-R charge radius, like the gravitational
radius, will then exhibit Regge behavior and
grow with an increasing energy of the probe.

Another important issue that needs to be addressed is that of
$p$-brane recoil. In the leading order calculations we have performed,
$p$-branes act as fixed objects which can absorb any amount of
transverse momentum. In the full quantum theory, however, we expect
a phenomenon similar to quantization of field theoretic collective
coordinates. As a result, the $0$-branes or the fully compactified
$p$-branes should recoil with velocity which depends on the soliton
mass. Some insight into this issue has recently been
obtained \cite{cf,willy}, but the Dirichlet $p$-brane formalism should
be helpful in performing detailed calculations.

\section*{Acknowledgements}

We are grateful to C. Callan, S. Gubser, J. Maldacena,
L. Susskind and E. Witten
for interesting discussions. We also thank Steve Corley for
pointing out that the discussion of the crosscap in the original
version of this paper was incorrect.
This work was supported in part by DOE grant DE-FG02-91ER40671,
the NSF Presidential Young Investigator Award PHY-9157482, and the
James S. McDonnell Foundation grant No. 91-48.

\end{document}